\documentclass[aps,prc,preprint,superscriptaddress,showpacs]{revtex4}
\usepackage{amsmath}
\usepackage{amssymb}
\usepackage{graphicx}
\usepackage{bm}
\newcommand{\be}{\begin{equation}}
\newcommand{\ee}{\end{equation}}

\newcommand{\bea}{\begin{eqnarray}}
\newcommand{\eea}{\end{eqnarray}}

\begin{document}



\title{SEMICLASSICAL DESCRIPTION OF AVERAGE PAIRING PROPERTIES IN NUCLEI
}

\author{X. VI\~NAS\textsuperscript{a}, 
P. SCHUCK\textsuperscript{b,c},
M. FARINE\textsuperscript{d}}
\address{
\textsuperscript{a} Departament d'Estructura i 
Constituents de la Mat\`eria
and Institut de Ci\`encies del Cosmos, \\
Facultat de F\'{\i}sica, Universitat de Barcelona,
Diagonal {\sl 647}, {\sl 08028} Barcelona, Spain\\
\textsuperscript{b}
Institut de Physique Nucl\'eaire,
IN2P3-CNRS,Universit\'e Paris-Sud,\\
{\sl 91406} Orsay-C\'edex, France\\
\textsuperscript{c} 
Laboratoire de Physique et Mod\'elisation des
Milieux Condens\'es,\\
CNRS and Universit\'e Joseph Fourier, Maison des Magist\`eres,\\
B.P. 166, 38042 Grenoble Cedex, France \\
\textsuperscript{d}
Ecole des Mines de Nantes, Universit\'e Nantes,\\
4, rue Alfred Kastler B.P. 20722\\
{\sl 44307}  Nantes-C\'edex 3, France
}
\begin{abstract}
\section{Abstract}
We present a new semiclassical theory for describing pairing in 
finite Fermi systems. It is based on taking the $\hbar \to 0$, i.e. 
Thomas-Fermi, limit of the gap equation written in the basis of the
mean field (weak coupling). In addition to the position dependence 
of the Fermi momentum, the size dependence of the matrix elements 
of the pairing force is 
also taken into account in this theory. An example typical for 
the nuclear situation shows the improvment of this new approach 
over the standard Local Density Approximation. We also show that 
if in this approach some shell fluctuations are introduced in the level 
density, the arch structure displayed by the quantal gaps 
along isotopic chains is almost recovered. We also point out that 
in heavy drip line nuclei pairing is strongly reduced.  
\end{abstract}
\maketitle
\pagebreak
\section{Introduction}

Averge properties in nuclei give very valuable information about 
salient features of nuclear physics. The most well known example is 
the celebrated Droplet Model and its extensions developed by Myers 
and \'{S}wi\c{a}tecki \cite{MS}. However, not only the average of binding 
energies is of interest. There are other properties like  
inertias \cite{far00,dur85}, one- and two-body matrix elements\cite{vin03}, 
etc. whose average behaviour is also of interest and has been 
studied in the past. Our aim here is to present a new semiclassical 
approach to nuclear pairing based on the Thomas-Fermi (TF) 
theory, i.e. in the $\hbar \to 0$ limit, to study the smooth 
behaviour of the pairing properties. Semiclassical approaches to 
the pairing problem of Finite Fermi systems may be helpful in cases where 
the quantal fluctuations are weak or if only the average behaviour is of 
interest.

The standard semiclassical limit to pairing is the well known Local 
Density Approximation (LDA) \cite{kuch89}. It consists in considering 
the BCS equations in infinite homogeneuous matter and replace the 
Fermi momentum $k_F$ by its local version in terms of the density.
LDA is valid in situations where the local Fermi wavelength $2 
\pi/k_F({\bf R})$ is small compared with the distance over which the 
mean field potential varies appreciably, that in the case of a 
harmonic oscillator (HO) potential $V({\bf R})=m \omega^2 R^2/2$ 
is the so-called  
oscillator length $l=\sqrt{\hbar/m \omega}$. Pairing introduces a 
second length scale that is the coherence length $\xi$ which gives 
the extension of the Cooper pairs. For LDA to be valid in the 
pairing case, also the condition $\xi/l << 1$ must be fulfilled. 
This condition is usually equivalent to $\Delta/\hbar \omega > 1$, 
where $\Delta$ is the gap in the single-particle spectrum. 
However, since in LDA $\xi \sim \Delta^{-1}$, there is a region in the tail 
of the surface where the condition $\xi/l << 1$ is violated because 
the gap vanishes in this region. In spite of these failures, 
integrated quantities as pairing energies may be quite accurate when 
considered in an average \cite{kuch89}.

We present here a semiclassical theory for pairing which 
improves the LDA. This theory shall be applicable in the weak pairing 
regime where $\mu \sim \varepsilon_F$ with $\mu$ the chemical 
potential and $\varepsilon_F$ the Fermi energy, and, therefore  
$\Delta/\mu << 1$. This TF theory works, for the average, in 
the region $\Delta < \hbar \omega$, where in general LDA breaks 
down. Preliminary applications of our theory has been made in 
\cite{far00,bar05}.

The contribution is organized as follows. The basic theory is 
sketched in the second section. Some results are presented in the 
third section. Our conclusions are given in the last section.
  
\section{Basic theory}

In the Hartree-Fock-Bogoliubov theory \cite{RS} the so-called canonical or natural 
basis, $|n_c \rangle$, diagonalizes simultaneously the single-particle density matrix 
$\hat \rho$ and the pairing tensor (anomalous density) matrix $\hat \kappa$:
\begin{equation}
\hat \rho|n_c \rangle = v_n^2|n_c \rangle, \quad 
\hat \kappa|n_c \rangle = u_nv_n|n_c \rangle,
\label{eq1}
\end{equation}
where the eigenvalues $v_n^2$ have the meaning of occupation numbers and the 
amplitudes $u_n, v_n$, normalised as $u_n^2 + v_n^2 = 1$,  are analogous to the ones 
also used in BCS theory \cite{RS}. 
In the canonical basis the gap equation has the following form:
\begin{equation}
\Delta_{n_c} =- \sum_{n'_c}V_{n_c n'_c}\frac{\Delta_{n'_c}}{2E_{n'_c}},
\label{eq2}
\end{equation}
where $V_{n_c n'_c} = \langle  n_c {\bar n}_c|v|n'_c {\bar n'}_c \rangle$   
is the matrix element of the pairing interaction with $|{\bar n}_c\rangle$ the
time reversed state of $|n_c\rangle$ and
$E_{n_c} = [(\epsilon_{n_c} - \mu)^2 + \Delta_{n_c}^2]^{1/2}$, with $\epsilon_{n_c}$
the diagonal elements of the normal mean field Hamiltonian and $\mu$ the chemical 
potential, are the eigenvalues of the HFB energy matrix \cite{RS}.

In the weak coupling regime we have $\Delta/{\mu} << 1$ and in this case the density 
and the density matrix are very little influenced by the
gap \cite{RS}. Therefore, one can replace with only little error the 
canonical
basis by the basis of the normal, non superfluid, mean field (H.F.)
Hamiltonian, that is $H|n \rangle = \epsilon_n|n \rangle$, which in terms of the 
density matrix $\hat \rho_n = |n\rangle \langle n|$ corresponding to the 
state $|n\rangle$ reads:
\begin{equation}
(H - \epsilon_n)\hat \rho_n = 0.
\label{eq3}
\end{equation}
We, thus, can write $\Delta_n = Tr[\hat \Delta \hat \rho_n]$ and
$\epsilon_n = Tr[H \hat \rho_n]$ and therefore
the state dependence of the gap equation (\ref{eq2}) is
entirely expressed through the density matrix $\hat \rho_n$.

Performing the Wigner transform (WT) of Eq.(\ref{eq3}) and using the fact that 
the WT of a product of two single-particle operators is, to lowest order in 
$\hbar$, equal to the product $A({\bf R},{\bf p})B({\bf R},{\bf p})$ of the 
individual WT, we  obtain for (\ref{eq3}) in the $\hbar \to 0$ limit\cite{RS}
\begin{equation}
(H_{cl.} - \epsilon)f_{\epsilon}({\bf R},{\bf p}) =0,
\label{eq5}
\end{equation}
where $f_{\epsilon}({\bf R},{\bf p})$ is the Wigner transform of $\hat \rho_n$
and $H_{cl.} = \frac{p^2}{2m^*({\bf R})} + V({\bf R})$ is the classical
Hamiltonian which contains a local
mean field potential $V({\bf R})$ and a position dependent
effective mass $m^*({\bf R})$. Eq.(\ref{eq5}) has to be unsderstood in the sense 
of distributions, therefore with $x\delta(x)=0$ we obtain
\begin{equation}
f_E({\bf R},{\bf p}) = \frac{1}{g^{TF}(E)}\delta(E - H_{cl.}) + O(\hbar^2),
\label{eq6}
\end{equation}
which is just the TF approximation to the normalized on-shell 
density matrix \cite{vin03}. The norm is equal to the level density (without 
spin-isospin degeneracy):
\begin{equation}
g^{TF}(E) = \frac{1}{(2\pi \hbar)^3} \int  d {\bf R} d {\bf p} \delta(E -
H_{cl.}).
\label{eq7}
\end{equation}

To derive the TF gap equation we replace $\hat \rho_n$ 
by its semiclassical counterpart (\ref{eq6}) everywhere in (\ref{eq2})
obtaining
\begin{equation}
\Delta(E) = \int_0^{\infty} dE' g^{TF}(E') V(E,E') \kappa(E')
\frac{\Delta(E')}{2\sqrt{(E'-\mu)^2 + \Delta^2(E')}},
\label{eq8}
\end{equation}
where the semiclassical pairing matrix element is written as \cite{vin03}
\begin{equation}
V(E,E') =
\int \frac{d {\bf R} d{\bf p}}{(2 \pi \hbar)^3}
\int \frac{d{\bf R'} d{\bf p'}}{(2 \pi \hbar)^3}
f_E({\bf R},{\bf p}) f_{E'}({\bf R'},{\bf p'})
v({\bf R},{\bf p};{\bf R'},{\bf p'}),
\label{eq10}
\end{equation}
with $v({\bf R},{\bf p};{\bf R'},{\bf p'})$ the double WT of
$<{\bf {r_1}} {\bf {r_2}} \vert v \vert {\bf {r_1'}} {\bf {r_2'}}>$ what
for a local translationally invariant force yields
$v({\bf R},{\bf p};{\bf R'},{\bf p'})=\delta({\bf R}-{\bf R'})v({\bf
p}-{\bf p'})$ where $v({\bf p}-{\bf p'})$ is the Fourier transform of the force
$v({\bf r}-{\bf r'})$ in coordinate space. For a density dependent zero 
range force, this gives $v_0(\rho({\bf R}))\delta({\bf R}-{\bf R}')$, 
with $v_0(\rho)$ the density dependent pairing strength \cite{gar99}.
Eqs.(\ref{eq8})-(\ref{eq10}) can be easily solved for a given mean field 
and the chemical potential is fixed by the usual particle number condition.


\section{Results}
\begin{figure}
\begin{center}
\includegraphics[height=8cm,angle=-90]{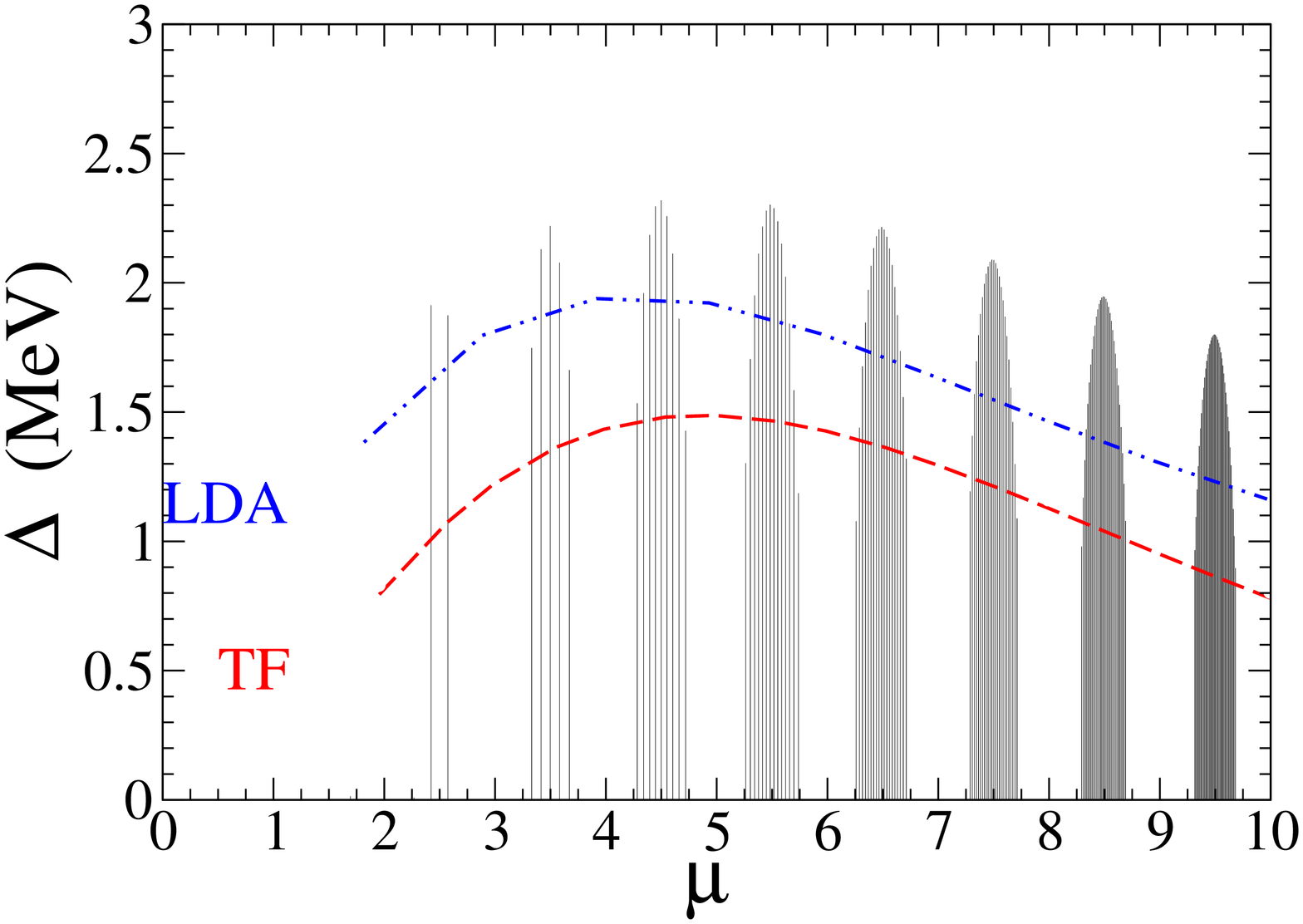}
\includegraphics[height=8cm,angle=-90]{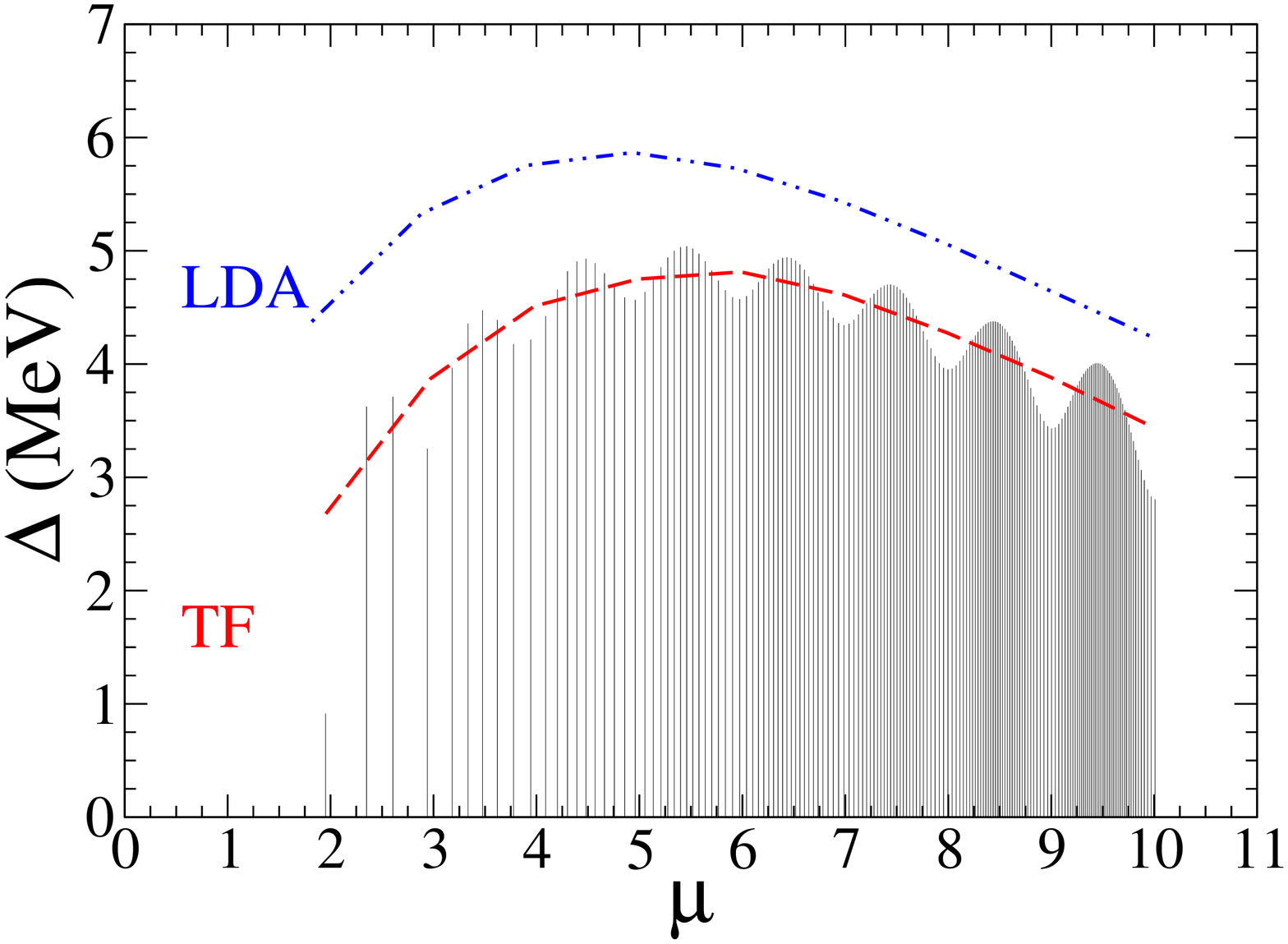}
\end{center}
\caption{\label{Fig1}Left: Quantal (solid line), TF (dashed line) 
and LDA (double dot-dashed) line pairing gap at the Fermi energy
as a function of the chemical potential in a HO potential.
Right: The same but with the strength of pairing force multiplied by 
a factor 1.5}
\end{figure}
First, we apply our TF theory for pairing to a model case.
To this end we will use the finite-range Gogny D1S force \cite{D1S}
for the pairing chanel together with a simplified mean field chosen as a 
simple HO potential well with constant $\hbar \omega$=8.65 
MeV and a constant effective mass $m^*/m$= 0.8. This model is 
schematic, since it does not include the spin-orbit potential, but this 
simplified description contains, however, the essential physics of the problem.

In the left panel of Figure 1 we display the gap at the Fermi energy as a 
function of the chemical potential $\mu$. In the quantal case we still perform an 
additional average over the substates in each major shell \cite{prak81}
that does not eliminate essential quantal features of the full solution, in 
particular the strong shell fluctuations are preserved. We see the break-down 
of pairing in the vicinity of closed HO shells. The dashed line 
represents our TF solution an the double dot-dashed curve shows the LDA 
values. We see that the TF approach averages the quantal values quite 
well while the LDA results seems to be too high. All three solutions
show globally a wide bump behavior. At small values of $\mu$, the level density
drops to zero and for high values of $\mu$ the matrix element
vanishes, since the Gogny force is finite range and cannot scatter 
the particles to very high energies. In between the two limits where the gap vanishes
, there is a 
maximum which has the same origine as the one of the gap in infinite matter
as a function of $k_F$ \cite{gar99}.
\begin{figure}
\begin{center}
\includegraphics[height=8cm,angle=-90]{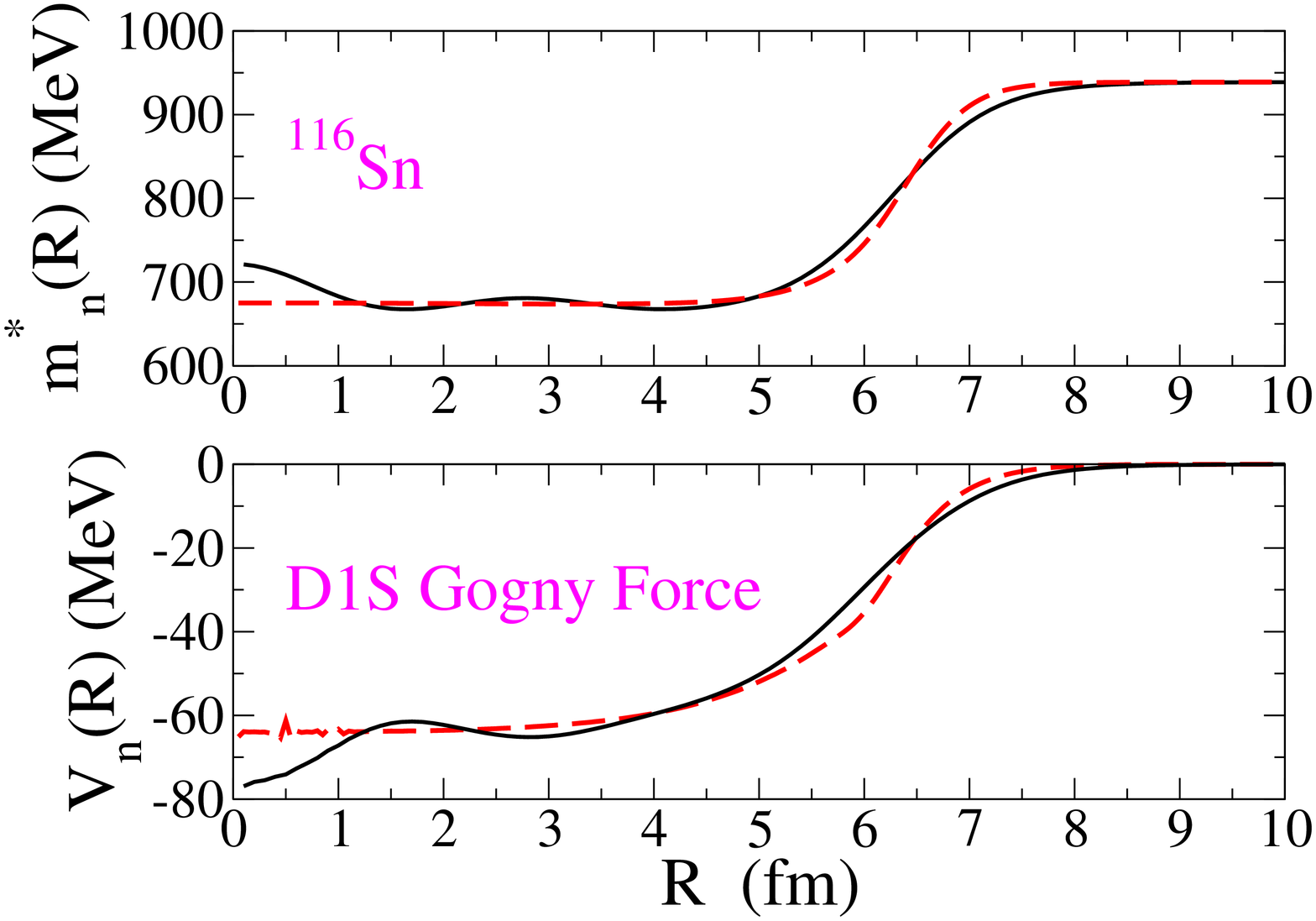}
\includegraphics[height=8cm,angle=-90]{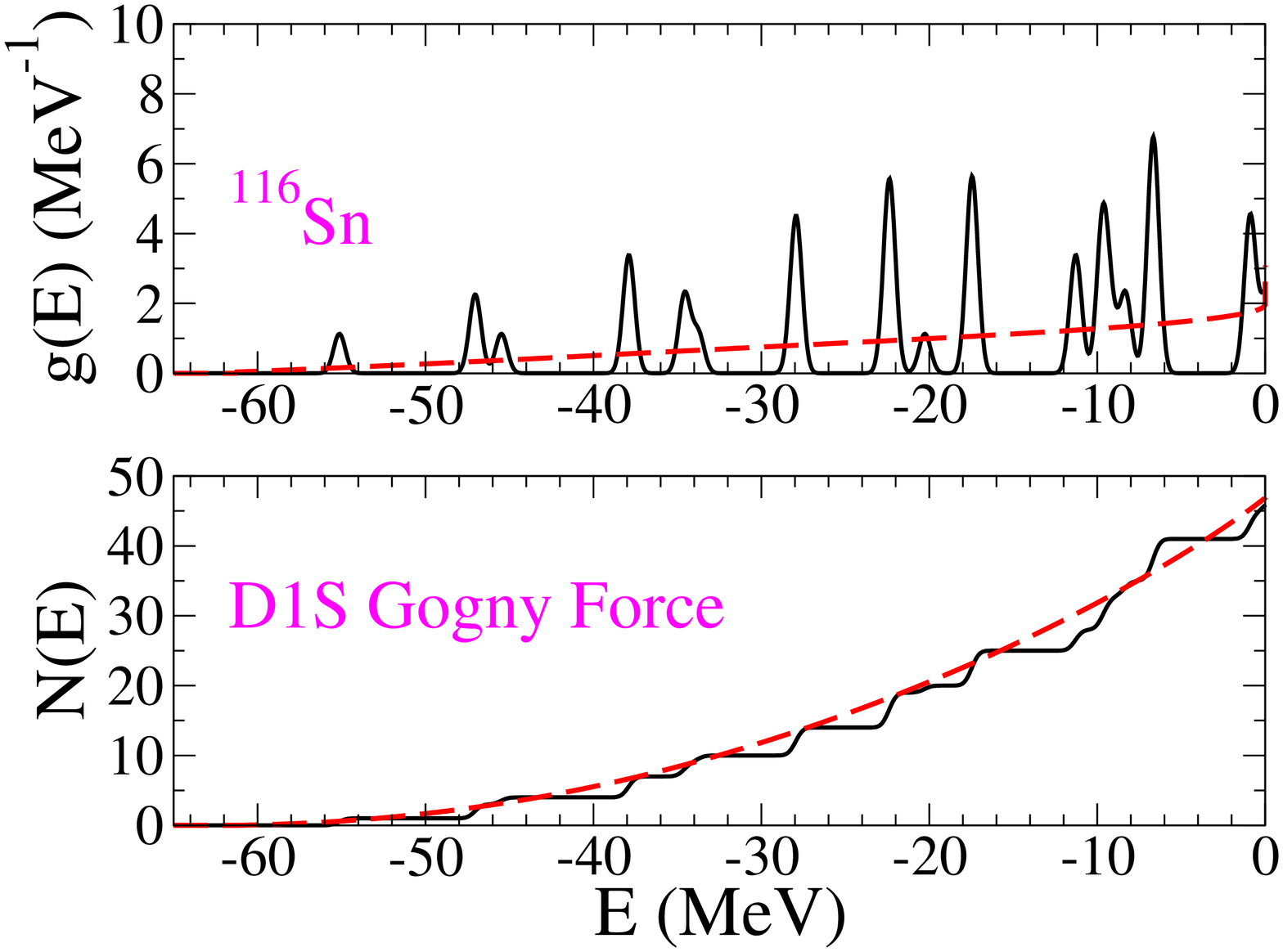}
\end{center}
\caption{\label{Fig2} Left: Neutron effective mass 
(top) and single-particle potential (bottom) for the nucleus
$^{116}$Sn. Right:Level density (top) and accumulated level density 
(bottom) of the nucleus $^{116}$Sn computed with fluctuations (solid 
line) and without (TF) (dashed line).} 
\end{figure}
In the right panel of Figure 1 we again show a similar study, only
the intensity of the pairing force has been increased by a factor of 1.5.
This increases the gap substantially and thus strongly smoothens the shell
effects of the quantal gaps. We see that in this scenario the TF approach
yields a precise average whereas LDA leads to an overestimation by about
20 percent. Increasing the force still further, the shell oscillations
disappear completely. We checked that in this case quantal and TF values
practically coincide whereas the LDA values remain too high.

As a realistic application of our TF theory for
pairing, we analyze the semiclassical pairing gaps as a function of mass 
number along the tin isotopic chain from $^{100}$Sn to $^{132}$Sn   
using the D1S Gogny force for both, mean field and pairing field.
The main ingredient for solving the semiclassical pairing equation 
(\ref{eq8}) is the on-shell density matrix $f_E({\bf R},{\bf p})$ 
(\ref{eq6}), which depends on the classical Hamiltonian $H_{cl}$ that is 
determined by the effective mass $m^*({\bf R})$  
and the mean field $V({\bf R})$.  
To obtain these quantities we use the Extended Thomas-Fermi (ETF) theory 
for finite-range non-relativistic interactions \cite{cent98,soub00}.
We compute $m^*({\bf R})$ and $V({\bf R})$ self-consistently and the 
corresponding semiclassical results are displayed by dashed lines 
in the left panel of Figure 2 in the case 
of the nucleus $^{116}$Sn. Using these quantities as input, one obtains 
the TF level density (\ref{eq7}) and the pairing matrix element (\ref{eq10}) 
which allow to solve the gap equation (\ref{eq8}) in our TF approximation. 

We can try to recover the arch structure by introducing additional quantal
information. As it is known for pairing, quantal fluctuations are  
more important in the level density than in the pairing matrix elements. 
To do that we proceed as follows. As it has been explained 
in Refs.\cite{soub00,soub03}, the ETF
energy density functional can be transformed, inspired by the 
Kohn-Sham scheme, into a quantal energy density functional. It should be 
noted that within this approximation the functional associated to a finite-
range effective interaction becomes local. The quantal $V({\bf R})$ and 
$m^*({\bf R})$  obtained in this way are also displayed by solid lines in 
Figure 2. We see, as expected, that the quantal oscillations are nicley 
averaged by the ETF solutions. 

For a given nucleus, once the single-particle energy levels have been
obtained, we build a fluctuating level density by taking a sum of
Gaussians each one centered at a single-particle energy, with a
width $\sigma=0.5$ MeV and with a strength such that the area below the
Gaussian equals the degeneracy of each energy level (spherical symmetry
is assumed). 
This fluctuating level density ${\tilde g}(E)$ corresponding to the nucleus
$^{116}$Sn is displayed in the top of the right panel of Figure 2, where we also show, for
comparison, the smooth TF level density $g^{TF}(E)$.
In the bottom of the same panel the accumulated level density with and without 
quantal fluctuations is also displayed.
We solve now the gap equation (\ref{eq8}) using the fluctuating level density 
${\tilde g}(E)$ but retaining the semiclassical matrix elements. 
The quantal features of pairing are essentially recovered in this way.
\begin{figure}
\begin{center}
\includegraphics[height=8cm,angle=-90]{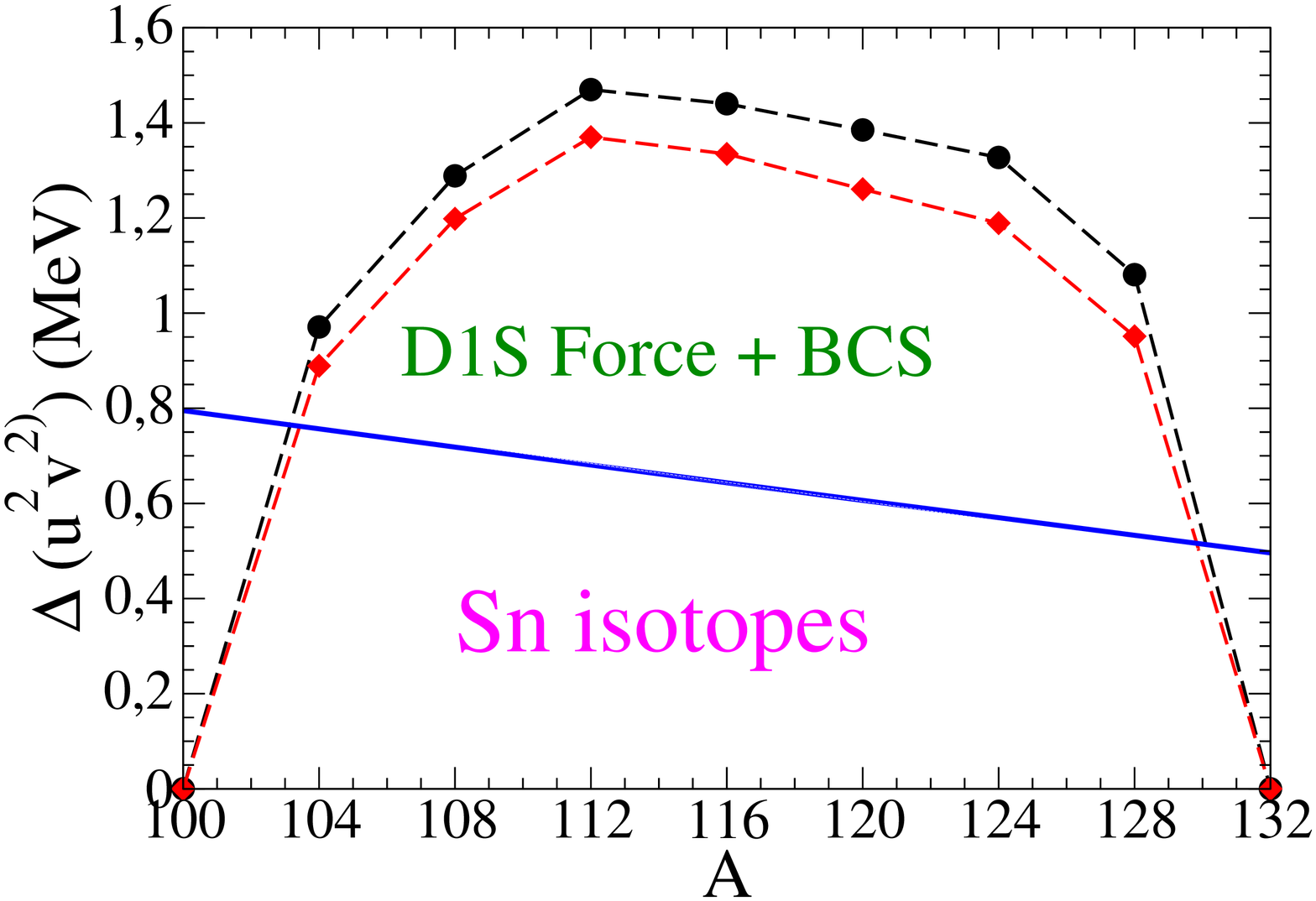}
\includegraphics[height=8cm,angle=-90]{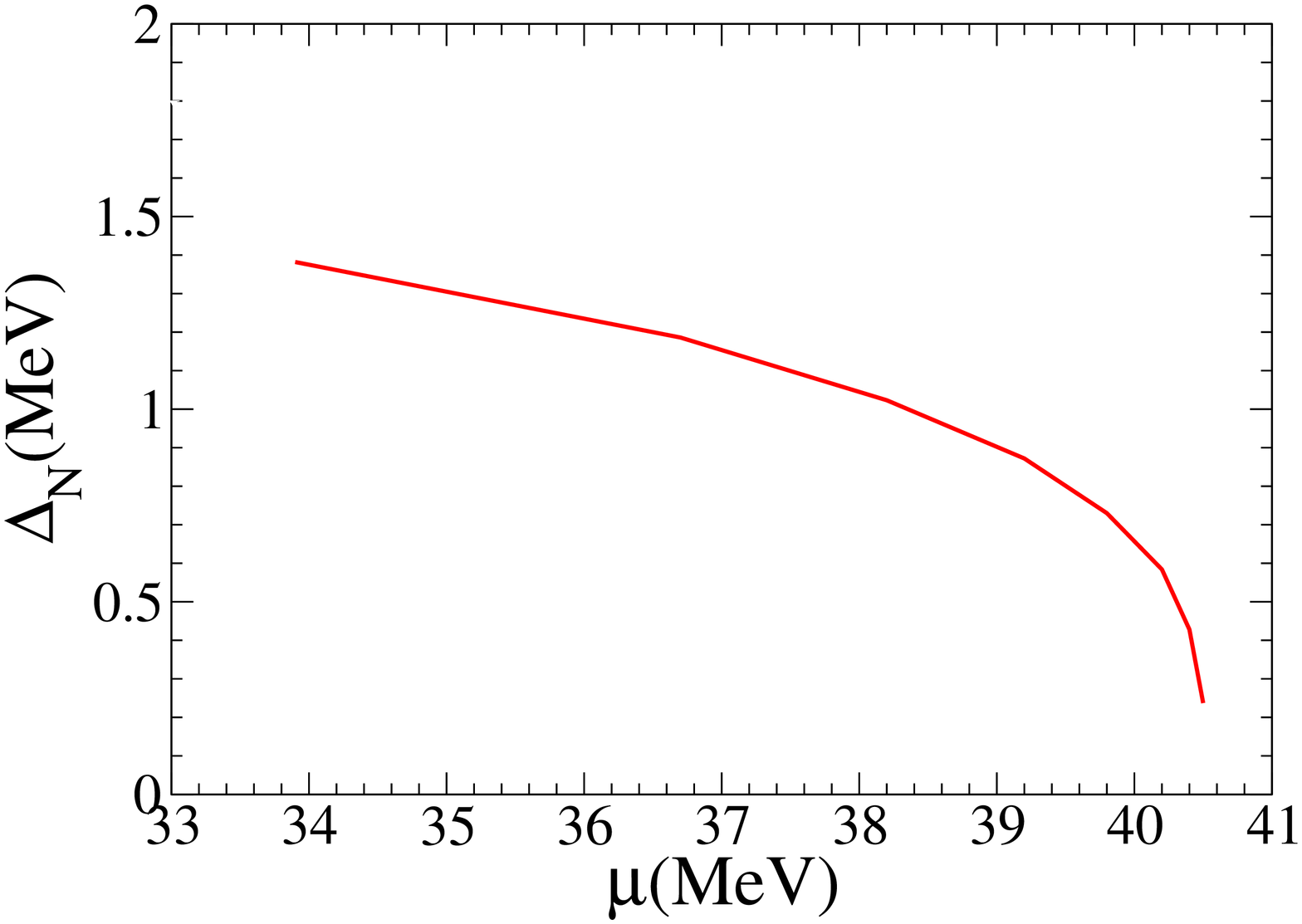}
\end{center}
\caption{\label{Fig3} Left: Average pairing gap along the Sn 
isotopic chain. See text for details. Right: Average TF gap at the 
Fermi energy as a function of the chemical potential
measured from the bottom of the single-particle potential and
computed in a box of $R$=100 fm.}
 \end{figure}
The average quantal gaps  
weighted with $u^2 v^2$ (circles) and the ones
obtained with our TF theory for pairing using the fluctuating
level density (diamonds) for the tin isotopic chain 
as a function of the mass number are displayed in Figure 3.
We see that by introducing ${\tilde g}(E)$ the
quantal arch structure is recovered and that the semiclassical
gaps obtained in this way reproduce quite accurately the quantal values.
In the same Figure we also display the semiclassical averages of the gap
computed with the smooth TF level density (solid line). We see that
in this case the quantal arch structure is completely washed out  
and that the semiclassical average gaps show a downward tendency with 
increasing neutron number. This downward trend also is clearly seen 
if one follows the quantal results over several major shells. 
In order to point out more clearly the behavior of the gap at the Fermi 
energy going to the drip line, we show in the right panel of the figure 
the behavior of the gap as a function of the chemical potential. 
We see that in the TF limit the gap vanishes when $\mu$
equals the depth of the single-particle potential, i.e. 
exactly at the drip. This should be relevant for instance for nuclei in 
the outer crust of neutron stars \cite{cha10} or for other finite Fermi systems 
with a large number of particles.
\section*{Conclusions}
We have presented a TF theory for pairing in finite Fermi
systems for weak coupling situations where $\Delta/\varepsilon_F << 1$.
This TF theory differs from the
usual LDA. This essentially stems from the fact
that we approximate the gap equation in configuration space and, thus,
keep the size dependence of the matrix elements of the pairing force. This is
not the case in LDA where the matrix elements of the force are always
evaluated in plane wave basis.
This semiclassical approach to pairing is only based on the usual validity
criterion of the TF theory, namely that the Fermi wave length is smaller
than the oscillator length. At no point the LDA
condition that the coherence length must be
smaller than the oscillator
length enters in the theory. Thus, the present TF approach yields for
all pairing quantities the same quality as TF theory does for
quantities in the normal fluid state.
An interesting feature of our study is that the average gap breaks down going
to the drip line. This unexpected result is confirmed by quantal calculations,
though strongly masked by shell fluctuations. For systems with large numbers 
of particles the fluctuations should die out and, thus, 
the semiclassical behaviour prevail.

\section*{Acknowledgments}

We thank Michael Urban for useful discussions and comments.
This work has been partially supported by the IN2P3-CAICYT collaboration. 
One of us (X.V.) acknowledges grants FIS2008-01661 (Spain and FEDER),
2009SGR-1289 (Spain) and Consolider Ingenio Programme CSD2007-00042 for 
financial support.

\end{document}